\documentclass[english]{article}
\usepackage[T1]{fontenc}
\usepackage[latin9]{inputenc}
\usepackage{geometry}
\usepackage{array}
\usepackage{textcomp}
\usepackage{multirow}
\usepackage{graphicx}

\makeatletter

\providecommand{\tabularnewline}{\\}

\newcommand{\lyxaddress}[1]{
\par {\raggedright #1
\vspace{1.4em}
\noindent\par}
}

\makeatother

\usepackage{babel}
\begin{document}

\title{Assessing influence of some meteorological parameters on airborne
particulate matters in the Himalayan hill-station of Darjeeling -
A preliminary assessment}

\author{Subhasis Banerjee{*} and Sanjay K. Ghosh}

\maketitle

\lyxaddress{Center for Astroparticle Physics and Space Science,, Bose Institute,
Kolkata, India}

\footnote{Disclaimer: This paper involves contribution from other scientists,
their names will appear in the final manuscript

{*}subhasis@jcbose.ac.in%
}

\section{Introduction}

Airborne particulate matters are know to pose serious threat to public
health in addition to its role in influencing the climate system.
In the past few decades efforts have been made to asses its level
in ambient air in different parts of the globe. Influence of several
atmospheric parameters on the concentration of particulate matters
in ambient air has also been studied in different parts of the world
{[}1-5{]}. Here we study influence of such few atmospheric parameters
on the level of RPM concentration in hill station of Darjeeling. We
performed correlation studies among RSPM concentration and selected
atmospheric parameters. We also performed stepwise multiple regression
analysis taking RSPM concentration as dependent variable to develop
statistical model. We analyzed the data taken for an entire year first
as a whole and then season wise.

\section{Data and methodology}

We used data collected in Darjeeling ( 27.0500\textdegree{} N, 88.2667\textdegree{}
E, average altitude form sea level 2050 meter), situated in the eastern
part of Himalaya in India, for a whole year (February 2012- January
2013). Atmospheric parameters have been provided by a automatic weather
station. The number of data points we have is sufficient only for
preliminary assessment. We have analyzed influence of atmospheric
parameters like surface temperature, atmospheric pressure, wind speed
and relative humidity on concentration of RSPM at first as whole and
then as per prominent seasons as witnessed in India. We divided our
data into different subsets according to seasons in order  to undertake
our study. We have only used those data points for which all the atmospheric
parameters are available.

\subsection*{2.1 Stepwise Regression}

Multiple linear regression(MLR) has been widely used by many authhors
in the field of air pollution. Multiple linear regression has the
form like
\[
Y=C_{0}+C_{1}X_{1}+C_{2}X_{2}+C_{3}X_{4}+....
\]
 Here, $Y$ is the dependent variable and $X_{i}$'s are predictor
variables. $C_{i}$'s are coefficients of regression. In stepwise
linear regression one performs simple MLR of all the predictor variables.
If all the predictor variables found significant then the total model
is good. Otherwise simple one variable linear regression is performed
with each of the predictor variables, and variable which gives lowest
p-value for t-test, is chosen. Then two-variable regression is performed
taking the chosen variable in the previous step as common. In this
way predictor variables are added in the model in each step (i.e.,
stepwise) as long as all the variables in the model are found significant. 

\begin{center}
\begin{table}
\begin{centering}
\begin{tabular}{|c|c|}
\hline 
Season & Month\tabularnewline
\hline 
Winter & December - February\tabularnewline
\hline 
Premonsoon & March - may\tabularnewline
\hline 
Monsoon & June - September\tabularnewline
\hline 
Postmonsoon & October - November\tabularnewline
\hline 
\end{tabular}
\par\end{centering}

\centering{}\caption{Monthwise prominent seasons in India}
\end{table}

\par\end{center}

\section{Results and Discussion}

In Table 1 we have compared RSPM concentration ($\mu g/m^{3}$) and
different atmospheric parameters among different seasons.
\begin{table}
\begin{centering}
\begin{tabular}{|c|c|c|c|c|c|c|c|c|c|c|}
\hline 
\multirow{2}{*}{Seasons/ Variables} & \multicolumn{2}{c|}{Overall } & \multicolumn{2}{c|}{Winter} & \multicolumn{2}{c|}{Pre-monsoon} & \multicolumn{2}{c|}{Monsoon} & \multicolumn{2}{c|}{Post-monsoon}\tabularnewline
\cline{2-11} 
 & Mean & SD & Mean & SD & Mean & SD & Mean & SD & Mean & SD\tabularnewline
\hline 
RSPM  & 33.88 & 22.07 & 45.54 & 22.11 & 41.02 & 21.08 & 20.77 & 18.59 & 25.70 & 15.41\tabularnewline
\hline 
Temperature & 11.50 & 4.59 & 5.77 & 1.83 & 13.13 & 3.02 & 16.40 & 0.91 & 10.01 & 2.45\tabularnewline
\hline 
Relative Humidity & 80.71 & 36.15 & 75.34 & 15.20 & 74.74 & 14.08 & 92.83 & 3.18 & 78.19 & 9.44\tabularnewline
\hline 
Atmospheric Pressure & 781.42 & 13.62 & 781.93 & 1.96 & 780.89 & 2.63 & 780.09 & 2.67 & 784.15 & 1.53\tabularnewline
\hline 
Wind Speed & 0.93 & 0.38 & 0.79 & 0.22 & 1.33 & 0.45 & 0.77 & 0.25 & 0.76 & 0.13\tabularnewline
\hline 
\end{tabular}
\par\end{centering}

\caption{Mean and standard deviation of RSPM concentration and some meteorological
parameters in different seasons in Darjeeling (SD stands for standard
deviation)}
\end{table}
 Observation shows during winter our study site recorded highest RSPM
concentration (Fig 1).
\begin{figure}
\includegraphics[angle=-90,scale=0.3]{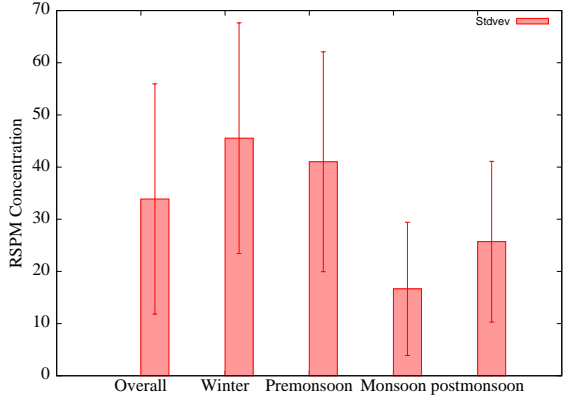}

\caption{Season wise concentration of RSPM. }
\end{figure}
 In Table 2 we presented Pearson's correlation co-efficient along
with p-value between atmospheric parameters and RSPM concentration
for different seasons.
\begin{table}
\begin{raggedright}
\begin{tabular}{|c|c|c|c|c|c|}
\hline 
Sesn. & Prmtr. & Corr. & P-val & Statistical Model & $R^{2}$\tabularnewline
\hline 
\multicolumn{1}{|c}{} & \multicolumn{4}{c}{} & \tabularnewline
\hline 
\multirow{4}{*}{\rotatebox{90}{Overall}} & Temp & -0.45 & 0.00 & \multirow{4}{*}{RSPM=1782.60 - 2.43 x Temp - 2.2 x Atmos\_prsr } & \multirow{4}{*}{0.26}\tabularnewline
\cline{2-4} 
 & RH & -0.04 & 0.73 &  & \tabularnewline
\cline{2-4} 
 & Atmos. Prsr. & -0.16 & 0.13 &  & \tabularnewline
\cline{2-4} 
 & wind Spd & 0.09 & 0.42 &  & \tabularnewline
\hline 
\multicolumn{6}{|c|}{}\tabularnewline
\hline 
\multirow{4}{*}{\rotatebox{90}{Winter}} & Temp & 0.14 & 0.51 & \multirow{4}{*}{RSPM=2892.68 + 0.9831 x RH - 3.76 x Atmos\_prsr + 3.74 x Temp} & \multirow{4}{*}{0.67}\tabularnewline
\cline{2-4} 
 & RH & 0.71 & 0.00 &  & \tabularnewline
\cline{2-4} 
 & Atmos Prsr & -0.44 & 0.04 &  & \tabularnewline
\cline{2-4} 
 & Wnd Spd & -0.10 & 0.64 &  & \tabularnewline
\hline 
\multicolumn{6}{|c|}{}\tabularnewline
\hline 
\multirow{4}{*}{\rotatebox{90}{Pre-} \rotatebox{90}{monsoon}} & Temp & -0.41 & 0.04 & \multirow{4}{*}{RSPM=74.80 - 2.63 x Temp} & \multirow{4}{*}{0.17}\tabularnewline
\cline{2-4} 
 & RH & -0.09 & 0.65 &  & \tabularnewline
\cline{2-4} 
 & Atmos Prsr & 0.08 & 0.72 &  & \tabularnewline
\cline{2-4} 
 & Wnd Spd & -0.17 & 0.39 &  & \tabularnewline
\hline 
\multicolumn{6}{|c|}{}\tabularnewline
\hline 
\multirow{4}{*}{ \rotatebox{90}{Monsoon}} & Temp & 0.10 & 0.66 & \multirow{4}{*}{None of the parameters found significant} & \multirow{4}{*}{}\tabularnewline
\cline{2-4} 
 & RH & 0.16 & 0.44 &  & \tabularnewline
\cline{2-4} 
 & Atmos Prsr & -0.41 & 0.04 &  & \tabularnewline
\cline{2-4} 
 & Wnd Spd & 0.15 & 0.49 &  & \tabularnewline
\hline 
\multicolumn{6}{|c|}{}\tabularnewline
\hline 
\multirow{4}{*}{\rotatebox{90}{Post-} \rotatebox{90}{monsoon}} & Temp & -0.59 & 0.01 & \multirow{4}{*}{RSPM=4536.24 - 68.35 x Wnd\_spd - 5.69 x Atmos\_prsr} & \multirow{4}{*}{0.69}\tabularnewline
\cline{2-4} 
 & RH & 0.44 & 0.08 &  & \tabularnewline
\cline{2-4} 
 & Atmos Prsr & -0.59 & 0.01 &  & \tabularnewline
\cline{2-4} 
 & Wnd Spd & -0.51 & 0.04 &  & \tabularnewline
\hline 
\end{tabular}
\par\end{raggedright}

\caption{Correlation (Pearson's) among different atmospheric parameters and
RSPM concentration and statistical model as derived by stepwise multiple
regression method with $R^{2}$.}
\end{table}
 We also presented the statistical model as derived by stepwise multiple
regression method. Our result clearly shows that in our study site
temperature and atmospheric pressure are strong negatively correlated
($p<0.05$) with RSPM concentration in most of the seasons. However
wind speed and relative humidity show negative and positive correlation
respectively with RSPM concentration for just one season only. Our
statistical model performs very well during winter and post-monsoon
seasons yielding $R^{2}$ value (coefficient of determination) 0.67
and 0.69 respectively indicating almost seventy percent of the variability
in RSPM concentration can be attributed to several atmospheric parameters.
In the season of monsoon none of the atmospheric parameters is significantly
correlated to RSPM concentration. As declared in the beginning, given
the size of the data sample, this study should be considered as preliminary
investigation however our efforts are on towards further detailed
study in this direction.

\end{document}